\documentclass[preprint,12pt]{elsarticle}

\newcommand{\beqn}{\begin{eqnarray}}
\newcommand{\eeqn}{\end{eqnarray}}
\newcommand{\eq}[1]{(\ref{#1})}

\newcommand{\Tr}{ {\rm Tr} \, }
\newcommand{\tr}{ {\rm Tr} \, }

\newcommand{\sign}{ {\rm sign} \,  }

\newcommand{\fm}{\mbox{fm}}
\newcommand{\Mev}{\mbox{MeV}}
\newcommand{\Gev}{\mbox{GeV}}

\newcommand{\plq}{\mbox{pl}}
\newcommand{\rt}{\mbox{rt}}

\usepackage{amsmath}
\usepackage{amstext}
\usepackage{amsfonts}
\usepackage{amssymb}
\usepackage{subfigure}
\usepackage{color}
\usepackage{epsfig}
\usepackage{bbm}
\newcommand{\spec}{\mbox{spec}}

\begin{document}

\begin{frontmatter}



\title{Magnetic polarizability of pion}


\author{E.V. Luschevskaya}
\ead{luschevskaya@itep.ru}
 \address{Institute for Theoretical and Experimental Physics, 117218 Russia, Moscow, Bolshaia Cheremushkinskaia 25\\
School of Biomedicine, Far Eastern Federal University, 690950 Vladivostok, Russia\\
National Research Nuclear University “MEPhI” (Moscow Engineering
Physics Institute), 115409, Russia, Moscow, Kashirskoe highway, 31}

  \author{O.E. Solovjeva}
\ead{olga.solovjeva@itep.ru}
\address{Institute for Theoretical and Experimental Physics, 117218 Russia, Moscow, Bolshaia Cheremushkinskaia 25}

\author{O.V. Teryaev}
\ead{teryaev@theor.jinr.ru}
 \address{Joint Institute for Nuclear Research, Dubna, 141980, Russia\\
National Research Nuclear University “MEPhI” (Moscow Engineering
Physics Institute), Kashirskoe highway 31, 115409 Moscow, Russia}

\begin{abstract}
We explore the energy dependence of   $\pi$ mesons off the
background   Abelian magnetic field on the base of quenched  SU(3)
lattice gauge theory and  calculate the magnetic dipole
polarizability of charged and neutral pions for various lattice
volumes and lattice spacings.  The contribution of the magnetic
hyperpolarizability to the neutral  pion energy has been also found.
\end{abstract}


\end{frontmatter}

 \section{Introduction}
\label{intro}

 Quantum Chromodynamics in strong magnetic fields is a promising
topic for research. Magnetic fields of hadronic scale could exist in
the Early Universe \cite{Hector:00} and could be formed in  cosmic
objects like magnetars and neutron stars. They can also be achieved
in terrestrial laboratories (RHIC, LHC, FAIR, NICA)
\cite{Skokov:2009}. The external electromagnetic fields can be
utilized as a probe of QCD properties, the recent progress obtained
on this way  in lattice gauge theories is discussed in
\cite{Massimo:2015}. The energy levels of hadrons in external
magnetic field   can be usefull for the calculation of  the  cross
sections \cite{Savage:2015}. The energies of mesons at the nonzero
magnetic field were calculated in various phenomenological
approaches
\cite{Simonov:2013,Liu:2015,Taya:2015,Kawaguchi:2016,Hattori:2016},
within the QCD sum rules \cite{Cho:2015,Gubler:2015} and in the
lattice gauge theories
\cite{Bali:2015,Luschevskaya:2015a,Luschevskaya:2015b}.

 Background magnetic Abelian fields also enable to calculate the
magnetic polarizabilities of  hadrons. In order to get the dipole
magnetic polarizability and the hyperpolarizability we measure the
energy of a meson as a function of the   magnetic  field. The
magnetic polarizabilities are important physical characteristics
describing the distibution of  quark currents inside a hadron  in an
external field. For the first time  the concept of polarizability
for the nuclear matter  was used by A.B. Migdal  in the analysis of
the scattering of low energy gamma quanta by atomic nuclei
\cite{Migdal}. For hadrons the notion of the polarizability was
discussed  in papers \cite{klein,baldin}.

 There were some discrepancies between the experimental obtained
value of the magnetic and the electric polarizabilities of the
charged $\pi$ mesons and some theoretical predictions based on the
chiral perturbation theory \cite{Gasser:2005,Aleksejevs:2013}.

Measurement of the electrical and the magnetic polarizabilities of
$\pi$ mesons was performed on the spectrometer SIGMA-AJAX in
Serpukhov, on the electron synchrotron Pakhra (LPI) in Moscow, on
the MarkII detector at SLAC, at COMPASS (CERN) and other
experiments.

According to the obtained data from these experiments, the value of
the polarizability of the charged $\pi$ mesons  is positive. The
most precise value of the charged pion electric polarisability has
been obtained experimentally  by the COMPASS     experiment
$\alpha_\pi= (2.0\pm 0.6_{stat}\pm 0.7_{sist})\times 10^{-4}\,
\fm^3$ \cite{Adolph:2015} under the assumption
$\alpha_\pi=-\beta_{\pi}$.  Comparison with the successful
predictions of the experiments and the chiral perturbation theory
\cite{Gasser:2005,Aleksejevs:2013} is interesting for fundamental
science.

In this work we consider a behaviour of the   ground state energy of
pions in the strong magnetic fields on the base of the SU(3) lattice
gauge theory. The details of the calculations are briefly sketched
in section  \ref{sec-2}. We  discuss the dipole magnetic
polarizabilities   of   $\pi^{\pm}$ and $\pi^0$ mesons in sections
\ref{sec-3} and \ref{sec-4} accordingly. The magnetic
hyperpolarizability of the neutral pion is calculated in   section
\ref{sec-5}.

\section{Details of calculations}
\label{sec-2}

\subsection{The improved gauge action}

\begin{table}[t]
\begin{center}
\begin{tabular}{|c|r|r|r|r|}
\hline
Ensemble & $ N_t \times N_s^3 $ & $ \beta_{imp}$  &$a,\ \fm$ & $ N_{conf} $  \\
\hline
$A_{16}$ & $16^4$          &      8.20       & 0.115    & 245       \\
\hline
$A_{18}$& $18^4$          &      8.10       & 0.125    & 285        \\

$B_{18}$ & $18^4$          &      8.20       & 0.115    & 200         \\
$C_{18}$& $18^4$          &      8.30       & 0.105    & 235          \\
$D_{18}$& $18^4$          &      8.45       & 0.095    & 195          \\
$E_{18}$& $18^4$          &      8.60       & 0.084    & 180          \\
\hline
$A_{20}$& $20^4$          &      8.20       & 0.115    & 275        \\

 \hline
\end{tabular}
\caption{The lattice simulations details. The  lattice volume is
shown in the second column, the lattice spacing and number of
configurations  used are presented in the fourth and the firth
columns  respectively.} \label{tbl1}
\end{center}
\end{table}

 For the generation of   quenched $SU(3)$  lattice configurations we
used the tadpole improved L\"uscher-Weisz action
\cite{Luscher:1985}, which reduces the ultraviolet lattice
artifacts. The action  has the form
\begin{equation}
S=\beta_{imp} \sum_{\plq} S_{\plq}-\frac{\beta_{imp}}{20
u^2_0}\sum_{\rt}S_{\rt}, \label{action}
\end{equation}
where $S_{\plq,\rt}= (1/2)\Tr(1-U_{\plq,\rt})$ is the plaquette
(denoted by $\plq$) or 1$\times$2 rectangular loop term ($\rt$),
$u_0=(W_{1\times1})^{1/4}=\langle(1/2)\Tr U_{\plq}\rangle^{1/4}$ is
the input tadpole factor computed at zero temperature
\cite{Bornyakov:2005}. Our simulations have been carried out on the
symmetrical lattices. The parameters of the lattice ensembles and
number of the configurations  are listed in the Table \ref{tbl1}.

\subsection{Fermionic spectrum}

We  solve  the Dirac equation numerically
\begin{equation}
D \psi_k=i \lambda_k \psi_k, \  \ D=\gamma^{\mu}
(\partial_{\mu}-iA_{\mu}) \label{Dirac}
\end{equation}
and found the eigenfunctions $\psi_k$ and the eigenvalues
$\lambda_k$ for a test quark in an external gauge field $A_{\mu}$.

For this goal we use the massive   overlap operator
~\cite{Neuberger:1997}. It  has the following form
 \begin{equation}
 M_{ov}=\left(1-\frac{am_q}{2 \rho}\right) D_{ov}+m_q,
 \label{massive_overlap_op}
 \end{equation}
where $m_q$ is the quark mass, $\rho=1.4$ is the  parameter in our
calculations, $D_{ov}$ is the massless overlap Dirac operator, which
preserves chiral invariance even at the finite lattice spacing $a$.
It may be written as
\begin{equation}
D_{ov}=\frac{\rho}{a}\left( 1+ \frac{D_W}{\sqrt{D_W^{\dagger} D_W}}
\right)=\frac{\rho}{a}(1+\gamma_5 \sign(H)), \label{overlap}
\end{equation}
where  $D_W=M-\rho/a$ is the Wilson-Dirac operator with the negative
mass term $\rho/a$,   $M$ is the   Wilson hopping term, $H=\gamma_5
D_W$ is the Hermitian Wilson-Dirac operator.

We construct the polynomial approximation for the  function
 \begin{equation}
\sign(H)=H/ \sqrt{H^{\dagger} H}. \label{sign_function}
\end{equation}
This approximation should be valid
  on the entire spectrum of $H$ matrix     $[\lambda_{min},\lambda_{max}]\in {\cal R}$.
Since
 \begin{equation}
 \sign(H)\equiv \sign(H/\Vert H \Vert)=\sign(W)
 \end{equation}
  and $\Vert H \Vert=\lambda_{max}$, then $\spec(W) \in [\lambda_{min}/\lambda_{max};1] \in {\cal R}$.
  The MinMax polynomial approximation is used for the function $1/\sqrt{H^2}$ at $\sqrt{\epsilon}\leq H < 1$, where $\epsilon=\lambda^2_{min}/\lambda^2_{max}$.

 The polynomial $P_n(H^2),\ H^2 \in [\epsilon;1]$  of a degree $n$ can be the best approximation, if it minimizes the maximal relative error
\begin{equation}
\delta = max |h(H^2)|
 \end{equation}
where $ h(H^2) =   1 - \sqrt{H^2}P_n(H^2) $.
 The polynom is represented by the series
\begin{equation}
P_n(H^2) = \sum_{k=0}^n c_k T_k(z),\ \ z =
\frac{2H^2-1-\epsilon}{1-\epsilon},
 \end{equation}
where $T_n(z),\, n=0,1,2,...$ are the Chebyshev polynomials defined
in the range $ [-1;1] $. The detailed description of the algorithm
used for the calculation of $P_n$   can be found in
\cite{Giusti:2003}.
   The resulting polynomial of the matrix  has the same
   eigenfunctions $\psi_k$ as the original matrix. The eigenvalues
   can be found from the eigenfunctions using  the formula
   $\lambda_k=\frac{\psi^{\dagger}_k|Q|\psi_k}{\psi^{\dagger}_k\psi_k}$
   for a some operator $Q$ \cite{Neff:2001}.
      From the polynomial approximation for the sign function we get the approximation for the overlap Dirac operator. We  find its eigenfunctions and eigenvalues, which are used for the calculation of the propagators and the correlators.

As we consider pure lattice gauge theory,   the Abelian magnetic
field is introduced   only into the overlap Dirac operator. In the
symmetric gauge  the magnetic field parallel to 'z' axis has the
form
  \begin{equation}
 A^B_{\mu} =\frac{B}{2} (x \delta_{\mu,2}-y\delta_{\mu,1}).
\end{equation}
 The total gauge field is the sum of the non-Abelian $SU(3)$ gluonic field and the Abelian $U(1)$
 field of magnetic photons
\begin{equation}
A_{\mu \, ij}= A_{\mu \, ij}^{gl} + A_{\mu}^{B} \delta_{ij},
\label{gaugefield}
\end{equation}
where $i,j=1,..N^2_c-1$ are the colour indices, $\mu=1,2,3,4$ are
the Lorentz indices.
 Quark fields obey periodic boundary conditions in space and
antiperiodic boundary conditions in time. In order to match
\eqref{gaugefield} with the periodic boundary conditions we apply
the additional x-dependent boundary twist for fermions
\cite{Al-Hashimi:2009}.

In the finite  lattice volume  the   magnetic flux   trough any
two-dimensional face of the hypecube is quantized. So the magnetic
field value is
\begin{equation}
eB=\frac{6\pi n_B}{(aN_s)^2}, \ \ n_B \in \mathbb{Z},
\label{quantization}
\end{equation}
where $e$ is the elementary charge and $N_s$ is the numbers of
lattice sites in spatial directions.

\subsection{Calculation of correlation functions}

To observe the ground state energy for  a meson we construct the
interpolating operator creating the state with the corresponding
quantum numbers. In case the pseudoscalar charged $\pi$  meson the
interpolating operator is described by the equations
\begin{equation}
{ O}(\pi^+)=\bar{\psi}_d (n) \gamma_5 \psi_u(n), \ \ \ {
O}(\pi^-)=\bar{\psi}_u (n) \gamma_5 \psi_d(n)
\end{equation}
The interpolating operator for the neutral pion is
\begin{equation}
{ O}(\pi^0)=(\bar{\psi}_u (n) \gamma_5 \psi_u(n)-\bar{\psi}_d (n)
\gamma_5 \psi_d(n))/\sqrt{2}.
\end{equation}

It should be mentioned that in Eucledean space
$\bar{\psi}=\psi^{\dagger}$. We are interested in the 2-point
lattice correlation function of the interpolating operators
\begin{equation}
\langle O_1 O_2\rangle_A=\langle\psi^{\dagger}(n) \Gamma_1 \psi(n)
\psi^{\dagger}(n^{\prime}) \Gamma_2 \psi(n^{\prime}) \rangle_A,
\label{observables}
\end{equation}
where  $\Gamma_1, \Gamma_2=\gamma_5,\, \gamma_{\mu}$ are Dirac gamma
matrices, $n=(\textbf{n} , n_t )$ and
$n^{\prime}=(\textbf{n}^{\prime}, n^{\prime}_t )$ are   lattice
coordinates. The spatial lattice coordinate
$\textbf{n},\textbf{n}^{\prime}\in
\Lambda_3=\{(n_1,n_2,n_3)|n_i=0,1,...,N-1\}$ and  $n_t,n^{\prime}_t$
are the numbers of lattice sites in  time direction.

The correlation function can be represented as the sum of connected
and disconnected parts
$$
\langle  O_1  O_2  \rangle_A=
-\tr[\Gamma_1D^{-1}(n,n^{\prime})\Gamma_2D^{-1}(n^{\prime},n)]
$$
\begin{equation}
 +\tr[\Gamma_1D^{-1}(n,n)]\tr[\Gamma_2D^{-1}(n^{\prime},n^{\prime})],
 \label{lattice:correlator}
\end{equation}
where $D^{-1}(n,n^{\prime})$ is the Dirac propagator. For the
isovector currents the disconnected part of the correlation function
has to be zero due to cancelation of $u$ and $d$ quarks
contributions, and we verify this on the lattice (see Sec.
\ref{sec-4}). The massive Dirac propagator is calculated using the
lowest $M=50$ Wilson-Dirac eigenmodes
\begin{equation}
D^{-1}(n,n^{\prime})=\sum_{k<M}\frac{\psi_k(n)
\psi^{\dagger}_k(n^{\prime})}{i \lambda_k+m}.
\label{lattice:propagator}
\end{equation}

 We  perform a discrete Fourier transformation of (\ref{lattice:correlator}) numerically
 and set $\langle\textbf{p}\rangle=0$ because  we are interested in the  ground state energy.
To obtain the masses we expand the correlation function to the
exponential series
 \begin{equation}
 \langle O_1 O_2 \rangle_A = \sum_k\langle 0|\hat{O}_1|k \rangle \langle k|\hat{O}^{\dagger}_{2}|0 \rangle e^{-n_t a E_k},
\label{sum}
 \end{equation}
where $\hat{O}$ and $\hat{O}^{\dagger}$ are the corresponding
Hilbert space operators.

For the large $n_t$ the main contribution to the correlator
(\ref{sum}) comes from  the term corresponding to the ground state.
Due to the periodic boundary conditions the correlator in the
leading order has the following form
\begin{equation}
\tilde{C}(n_t)= 2A_0 e^{-N_T a E_0/2} \cosh ((\frac{N_T}{2}-n_t) a
E_0),
 \label{sum33}
\end{equation}
where  $A_0$ is a constant, $E_0$ is the ground state energy,  $a$
is the lattice spacing. We found  the  ground state energies fitting
the lattice correlators by the equation \eq{sum33} at $5 \leq  n_t
\leq N_t-5$.

\section{Dipole  polarizability of   $\pi^{\pm}$ meson}
\label{sec-3}
\begin{figure}[htb]
\begin{center}
\includegraphics[width=7cm,angle=-90]{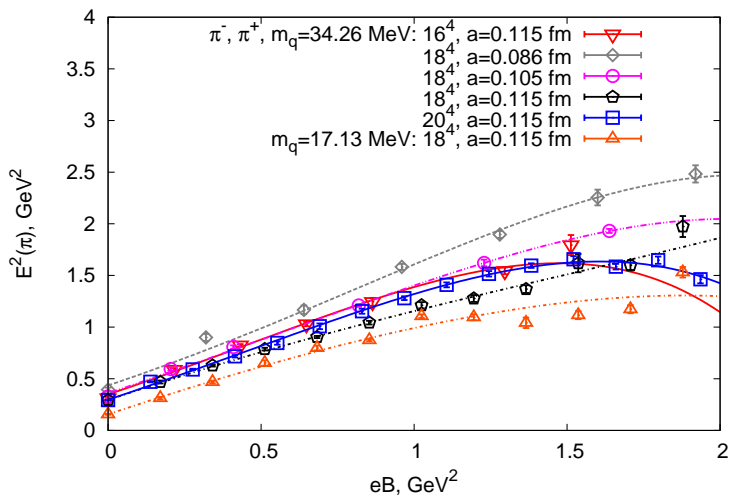}
\caption{The  qround state energy squared of $\pi^{\pm}$ meson
versus the  field  value squared for the lattice volumes $16^4$,
$18^4$, $20^4$, various lattice spacings and the quark masses
$17.13\ \Mev$ and $34.26\ \Mev$. Fits of the  lattice data
correspond to  theoretical formula \eq{eq4} at $eB\in [0, 2]\
\Gev^2$.} \label{fig-4}
\end{center}
\end{figure}

The energy levels of  a free charged  pointlike  particle in a
constant Abelian magnetic field parallel to z axis are described by
the following equation:
 \begin{equation}
E^2=p^2_z+(2n+1)|qH|-gs_zqH+E^2(H=0), \label{eq3}
\end{equation}
where  $p_z=0$ is the particle momentum along 'z' axis, $n$ is the
principal quantum number, $q$ is the particle charge,  $s_z$ is its
spin projection, $g$ is the dimensionless quantity, that
characterizes the magnetic moment of the particle and $E(H=0) \equiv
m$ is the energy at zero magnetic field (the mass).

For  the charged pion in the ground state in the rest frame  it is
necessary to utilize  $p_z=0$,    $n=0$,    $q=\pm 1$ and   $s_z=0$.
In   the strong external magnetic field the charged pion is not  a
pointlike particle anymore and its internal structure can be
described by the    magnetic polarizabilities. If we take into
account   the dipole magnetic polarizability $\beta_m$ and the first
order magnetic hyperpolarizability  $\beta^h_m$, then in the
relativistic case  the pion energy squared has the following form
  \begin{equation}
E^2=|qH| +m^2-4\pi m \beta_m H^2 - 4\pi m \beta^{1h}_m H^4, \ \
H=eB. \label{eq4}
\end{equation}

The energy of a charged pion was calculated from the correlation
function
\begin{equation}
C^{PSPS}=\langle \bar{\psi}_d(\vec{0}, n_t)\gamma_5 \psi_u(\vec{0},
n_t)  \bar{\psi}_u(\vec{0}, 0)\gamma_5 \psi_d(\vec{0}, 0) \rangle.
\end{equation}

 The energies squared  are  shown in Fig.\ref{fig-4}  for the
lattice volumes $16^4$, $18^4$, $20^4$, lattice spacings $0.086\
\fm$, $0.105\ \fm$, $0.115\ \fm$, quark masses $17.13\ \Mev$ and
$34.26\ \Mev$. We   fit our lattice data by formula \eq{eq4} at
$eB\in [0, 2]\ \Gev^2$, where $m$, $\beta_m$ and  $\beta^{1h}_m$ are
the fit parameters. The fitting curves are  also depicted in
Fig.\ref{fig-4}.  At the magnetic fields   $eB \lesssim 0.5\ \Gev^2$
we observe a linear dependence of the energy squared versus the
magnetic field $eB$ and it increases with the field for the all sets
of lattice data. At higher   fields the nonlinear terms in the
magnetic field  contribute to the pion energy.

Fitting the lattice data we do not consider the term with the second
order magnetic hyperpolarizability ($4\pi m \beta^{2h}_m H^6$) in
\eq{eq4}, because   from the data fits it follows that its relative
contribution   is small in comparison with the terms proportional to
$H^2$ and $H^4$.

In Table \ref{Table4} we show the values of the magnetic dipole
polarizability and hyperpolarizability for several lattice data
sets. The best values obtained are $\beta_m=(-1.15 \pm 0.31) \times
10^{-4}\ \fm^3$ for the lattice volume $20^4$, lattice spacing
$0.115\ \fm$ and  $\beta_m=(-2.06 \pm 0.76) \times 10^{-4}\ \fm^3$
for the lattice volume $18^4$ and lattice spacing $0.086\ \fm$.

In  two loops of the chiral perturbation theory it was predicted
that the   magnetic polarizability of $\pi^{\pm}$   equals to $-2.77
\times 10^{-4}\ \fm^3$ \cite{Aleksejevs:2013,Ivanov:2015}, that is
close to our result. The    polarizabilities for the $V=18^4$ and
$a=0.115\ \fm$ lattice at quark mass $m_q=34.26\ \Mev$ and
$m_q=17.13\ \Mev$   are not represented in   Table \ref{Table4}
because of the poor relative accuracy.  It is a very subtle effect
but  the work in this direction is carried out,  including the
smaller bare quark masses.

In 2015 the COMPASS  collaboration at CERN has investigated the pion
Compton scattering $\pi^- \gamma\rightarrow \pi^- \gamma$
\cite{Adolph:2015}. They have found the   pion electric
polarizability equal to $\alpha_{\pi^{\pm}}=(2.0\pm 0.6_{stat}\pm
0.7_{syst})\times 10^{-4}\ \fm^3$ under the assumption
$\alpha_{\pi^{\pm}}=-\beta_{\pi^{\pm}}$ of the ChPT, which is true
in the exact chiral limit \cite{Donoghue:1989}. We have observed
agreement of this value with our lattice results. This is also in
accordance with the earlier analysis of MARK II group data of the
cross section of the process $e^{+} e^{-}\rightarrow e^{+} e^{-}
\pi^{+} \pi^{-}$  \cite{Boyer:1990} which was made  in
\cite{Babusci:1992} and gave the value
$\alpha_{\pi^{\pm}}=-\beta_{\pi^{\pm}}=(2.2\pm
1.6_{stat+syst})\times 10^{-4}\ \fm^3$.

The Serpukhov group have found the value $\alpha_{\pi^{\pm}}=(6.8
\pm 1.4_{stat}\pm 1.2_{syst})\times 10^{-4}\ \fm^3$ exploring the
radiative pion nucleon scattering $\pi^{-} Z\rightarrow \pi^{-} Z
\gamma$ in 1983 \cite{Antipov:1983}. Without using the relation
$\alpha_{\pi^{\pm}}+\beta_{\pi^{\pm}}=0$ they obtained
$\alpha_{\pi^{\pm}}+\beta_{\pi^{\pm}}=(1.4\pm3.1_{stat}\pm2.5_{syst})\times
10^{-4}\ \fm^3 $.

The analysis of the  experimental data on helicity amplitudes of the
reaction $\gamma \gamma \rightarrow \pi^+ \pi^-$ gave the value
$\alpha_{\pi^{\pm}}= - \beta_{\pi^{\pm}}=6.41 \times 10^{-4}\ \fm^3$
\cite{Filkov:2006}.

\section{Dipole polarizability of   $\pi^0$ meson}
\label{sec-4}

\begin{figure}[htb]
\begin{center}
\includegraphics[width=7cm,angle=-90]{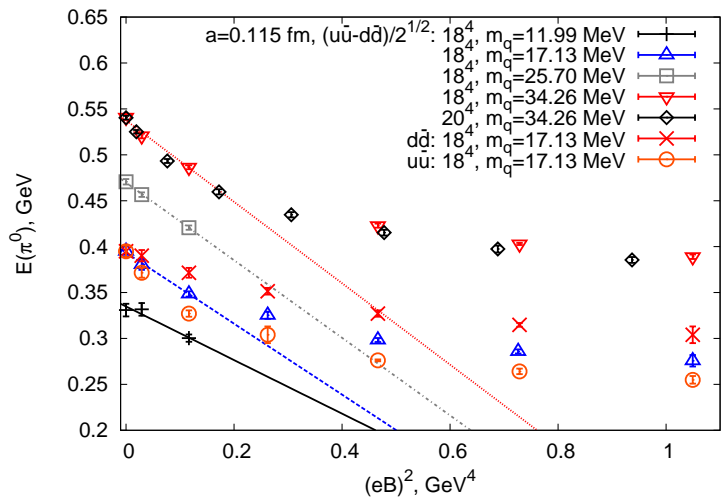}
\caption{The  ground state energy of $\pi^0$ meson  as a function of
the field value squared for various lattice volumes, spacings and
quark masses. Points correspond to the lattice data,  lines are  the
fits to the lattice data  obtained using function \eq{eq1}.}
\label{fig-1}
\end{center}
\end{figure}

 \begin{figure}[htb]
\begin{center}
\includegraphics[width=7cm,angle=-90]{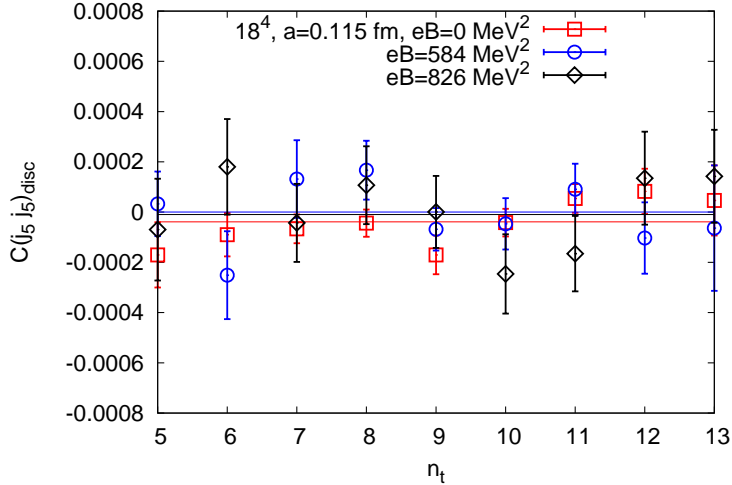}
\caption{The disconnected part of the correlator for neutral pion
for the lattice spacing $0.115\ \fm$, lattice volume $18^4$ and bare
quark mass $34.26\ \Mev$ for several values of the magnetic field.}
\label{fig:disc}
\end{center}
\end{figure}

\begin{figure}[htb]
\begin{center}
\includegraphics[width=7cm,angle=-90]{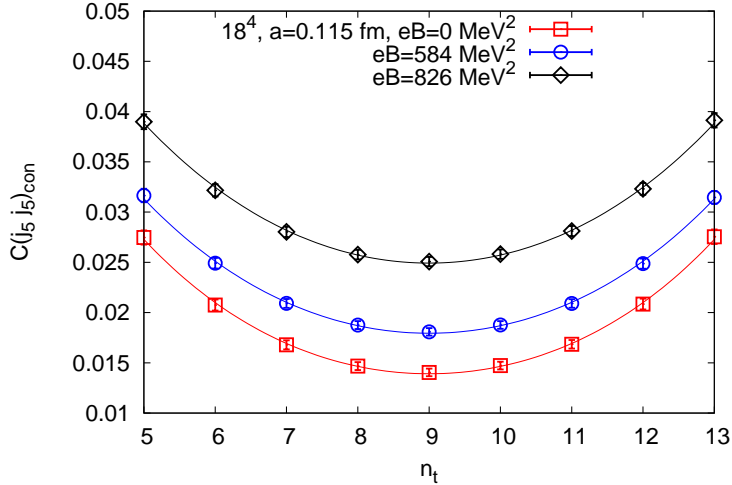}
\caption{The connected part of the $\pi^0$ correlator   for the
lattice spacing $0.115\ \fm$, lattice volume $18^4$ and bare quark
mass $34.26\ \Mev$ for several values of the magnetic field.}
\label{fig:con}
\end{center}
\end{figure}

\begin{figure}[htb]
\begin{center}
\includegraphics[width=7cm,angle=-90]{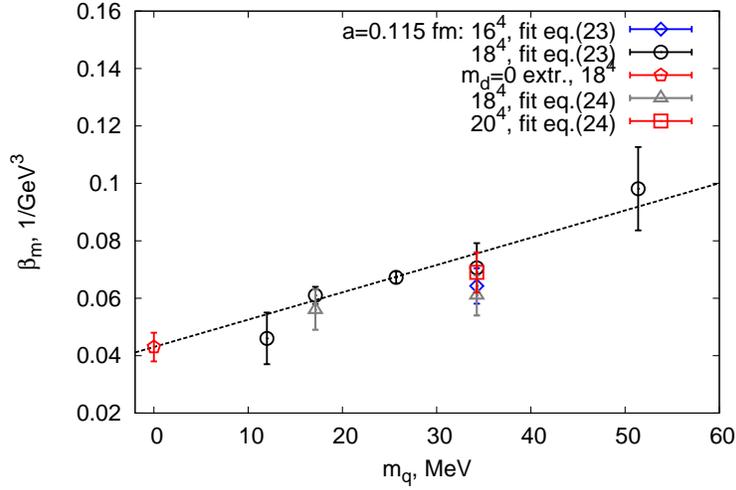}
\caption{The dipole magnetic  polarizability of     $\pi^0$ meson
for the   lattice volumes $16^4,\ 18^4,\ 20^4$ and lattice spacing
$0.115\ \fm$  depending on the quark mass.} \label{fig-2}
\end{center}
\end{figure}

We calculate the energy of neutral pion  from the correlation
function
$$C^{PSPS}=(\langle \bar{\psi}_d(\vec{0}, n_t)\gamma_5 \psi_d(\vec{0}, n_t)  \bar{\psi}_d(\vec{0}, 0)\gamma_5 \psi_d(\vec{0}, 0) \rangle$$
\begin{equation}
+\langle \bar{\psi}_u(\vec{0}, n_t)\gamma_5 \psi_u(\vec{0}, n_t)
\bar{\psi}_u(\vec{0}, 0)\gamma_5 \psi_u(\vec{0}, 0) \rangle)/2.
\end{equation}
It is taken into account that the   $d$ and $u$ quarks interact with
the external magnetic field differently because of their non-equal
charges.

  We have checked on the lattice that the disconnected part of the
 correlator \eqref{lattice:correlator} is zero within the error
 range. In Fig.\ref{fig:disc} it is shown for three values of the
 magnetic field.
In Fig.\ref{fig:con} the connected   correlators  are represented
with the fits \eqref{sum33} for comparison. In the case of exact
isospin symmetry the
 disconnected part has to be zero so in what follows we neglect  it at all.

 Fig.\ref{fig-1} shows the ground state  energy of the $\pi^0$ meson
depending on the   field value squared  for   ensembles  $A_{20}$ at
the quark mass $34.26\ \Mev$ and $B_{18}$ at the quark masses
$11.99\ \Mev,\ 17.13\ \Mev,\ 25.70 \ \Mev$, $34.26\ \Mev$ (see Table
\ref{tbl1}). If the neutral pion would consist  of one type of
quark, $d\bar{d}$ ($u\bar{u}$), then its energy decreases slower
(faster) than for the real pion, see Fig.\ref{fig-1}.

For the pion each lattice data set is  fitted at $(eB)^2 \in
[0,0.15\ \Gev^4]$  using the function
 \begin{equation}
E=E(B=0)- 2 \pi \beta_m(eB)^2, \label{eq1}
\end{equation}
where  we determine $E(B=0)$ and $\beta_m$ as the fit parameters.
$E(B=0)$  is  the energy at zero magnetic field  and $\beta_m$ is
the magnetic polarizability  of neutral pion which is  presented in
Table \ref{Table2}. We observe the linear energy dependence on the
magnetic field squared at  small values of the field.

In  Fig.\ref{fig-2} we represent the $\beta_m$ values of  the
$\pi^0$ meson     as a function of the quark mass  for  ensemble
$B_{18}$, the $\beta_m$ values for ensembles $A_{16}$, $A_{20}$ at
quark mass $34.26\ \Mev$ are also depicted.

The value of   $\beta_m(\pi^0)$ diminishes with the quark mass and
extrapolation to the chiral limit gives the number  $(3.3\pm
0.4)\cdot 10^{-4} \ \fm^3$  for the lattice volume $18^4$ and
lattice spacing $0.115\ \fm$. It is a   bit higher than the value of
magnetic polarizability obtained in our previous work
\cite{Luschevskaya:2015a}, because  we considered $\pi^0$ meson
consisting of one type of quarks  $d$.
 In the framework of the chiral perturbation theory the  value of the $\beta_m(\pi^0)$
 is equal to $\beta_m=0.5 (1.5\pm 0.3) \cdot 10^{-4}\ \fm^3$ in one (two) loops \cite{Gasser:2005}.

Our result coincides in sign with the prediction of the ChPT and
differs in value, and there can be various reasons for that,
including the contributions of higher chiral loops. In turn, from
the lattice side, the dynamical quark loops may be relevant.

\section{Magnetic hyperpolarizability of neutral pion}
 \label{sec-5}

 \begin{figure}[htb]
\begin{center}
\includegraphics[width=7cm,angle=-90]{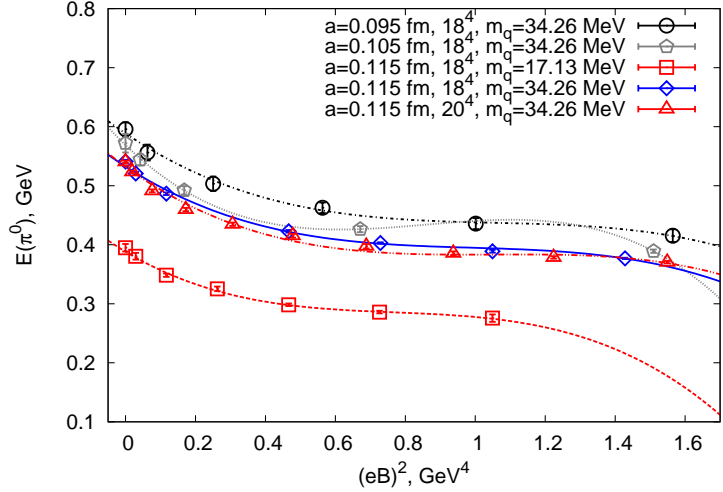}
\caption{The  ground state energy   of $\pi^0$ meson versus the
field  value squared for the lattice volumes $18^4$,  $20^4$, the
lattice spacings $0.095\ \fm,\ 0.105\ \fm,\ 0.115\ \fm$ and   quark
masses $17.13\ \Mev$, $34.26\ \Mev$. Fits of the  lattice data were
carried out using the euqation \eq{eq2} at $eB\in [0, 1.7]\
\Gev^2$.} \label{fig-3}
\end{center}
\end{figure}

The terms of the higher degrees of the magnetic field   give
contribution to the   energy of the pseudoscalar meson at very high
magnetic field. Fig.\ref{fig-3} shows the $\pi^0$ ground state
energy     depending on the   magnetic field squared at $eB \in
[0,1.7]\ \Gev^4$. The lattice data fits are described by the formula
 \begin{equation}
E=E(B=0)- 2 \pi \beta_m(eB)^2-2\pi \beta^h_m (eB)^4 -  k (eB)^6,
\label{eq2}
\end{equation}
where     $\beta^h_m$ is the magnetic  hyperpolarizability and $k$
is the hyperpolarizability of higher order,  $\beta_m$, $\beta^h_m$
and $k$ are the fit parameters. The term proportional to $B^3$ is
parity forbidden: the physical interpretation is that $\pi_0$ can't
decay to three (magnetic) photons.

 The decay  $\pi^0\rightarrow 4 \gamma_M$ is allowed, so the  term $\sim (eB)^4$ presents in formula  \eq{eq2}.
  The values $\beta^h_m$  are negative for the all sets of the lattice data and showed in Table \ref{Table3}.

  We  found the value $\beta^h_m= (-6.7\pm 1.7) \cdot 10^{-7} \ \fm^7$ at lattice spacing $a=0.095\ \fm$   and $\beta^h_m=   (-6.8\pm 1.5) \cdot 10^{-7} \ \fm^7$ at $a=0.115\ \fm$ for the lattice volume $18^4$, and quark mass $34.26\ \Mev$ .

  At the lowest quark mass $17.13\ \Mev$ we   obtained  $\beta^h_m(\pi^0)=   (-7.8 \pm 1.7) \cdot 10^{-7} \ \fm^7$   using for fitting  formula \eq{eq2}.

Note also that the energy shows the qualitative tendency of
flattening at large magnetic field so that it should not turn to
zero and cause the phase transition, in complete similarity to
$\rho-$meson case studied earlier \cite{Luschevskaya:2015b}. The
decrease at very large fields is likely to be compensated by higher
hyperpolarizabilities. Still, the quantitative analysis of this
behaviour requires more investigations.

\section{Conclusions}
 \label{sec-6}

We  have calculate  the magnetic dipole polarizability of the
charged pion for several lattice spacings,  lattice volumes $16^4,\
18^4,\ 20^4$ and  bare lattice quark mass $34.26\ \Mev$. For the
lattice volume $20^4$ and the lattice spacing $0.115\ \fm$ we obtain
the value $\beta_m=(-1.15 \pm 0.31)\cdot 10^{-4}\ \fm^3$. At finest
lattice with the lattice spacing $0.086\ \fm$ and the volume $18^4$
we get   $\beta_m=(-2.06\pm 0.76)\cdot 10^{-4}\ \fm^3$. We found the
agreement of this result  with the observation of COMPASS
collaboration.

We have explored the energy dependence of neutral pion off the
external Abelian magnetic field and have found its dipole magnetic
polarizability and hyperpolarizability. The dipole magnetic
polarizability depends on the lattice quark mass and the chiral
extrapolation was performed.
  In the chiral limit we have obtained the magnetic dipole
polarizability of the $\pi^0$ is equal to  $(3.3\pm 0.4)\cdot
10^{-4} \ \fm^3 $ for the lattice volume $18^4$ and the lattice
spacing $0.115\ \fm$, that is close to the prediction  of the ChPT
\cite{Aleksejevs:2013}.

The  contribution of the hyperpolarizability $\beta^h_m$ have been
also revealed   to the $\pi^0$ energy at considered strong magnetic
field. This is very tiny effect and  for the all sets $\beta^h_m<0$.
For the finest lattice with spacing $a=0.095\ \fm$ we found
$\beta^h_m= (-6.7\pm 1.7) \cdot 10^{-7} \ \fm^7$. We have observed
qualitative behaviour of energy which is not favouring the
appearance of tachyonic mode and phase transition.

\section{Acknowledgements}
O.T. is indebted to A. Guskov and M. Ivanov for useful discussions.

   The authors are grateful to  FAIR-ITEP supercomputer center where these numerical calculations were performed.
   This work   is completely supported by a grant from the Russian Science Foundation (project number 16-12-10059).

\begin{table}[htb]
 \begin{center}
\begin{tabular}{|c|r|r|r|r|r|}
\hline
$V_{latt}$  & $m_d$ $(\Mev)$ &$a\ (\fm)$   &$\beta_m\,(\Gev^{-3})$  & $\beta^{1h}_m\,(\Gev^{-7})$ & $ \chi^2$/d.o.f.  \\
\hline
$18^4$           & $34.26$  & $0.086$    &   $-0.027   \pm 0.010   $   & $0.007 \pm 0.004$ & $4.828$   \\
\hline
 $20^4$          & $34.26$  & $0.115$    &   $-0.015 \pm 0.004$        & $0.011 \pm 0.002$ & $5.121$  \\
\hline
 $18^4$          & $34.26$  & $0.105$    &   $-0.011 \pm 0.007$        & $0.005 \pm 0.003$ & $3.414 $   \\
 \hline
$16^4$           & $34.26$  & $0.115$    &   $-0.021 \pm 0.015 $       & $0.015 \pm 0.009$ &  $3.935$  \\
\hline
\end{tabular}
\end{center}
\caption{The magnetic dipole polarizability $\beta_m$ and
hyperpolarizability $\beta^h_m$ of $\pi^{\pm}$, its errors  and $
\chi^2$/d.o.f. of  fit \eq{eq4} to the data at $eB  \in [0,2]\
\Gev^2$ for the bare quark masses $m_q=34.26\ \Mev$, various lattice
volumes and spacings.} \label{Table4}
 \end{table}

\begin{table}[htb]
 \begin{center}
\begin{tabular}{|c|r|r|r|r|}
\hline
$V_{latt}$  & $m_d$ $(\Mev)$ &$a\ (\fm)$   &$\beta_m\,(\Gev^{-3})$  &  $ \chi^2$/d.o.f.  \\
\hline
$16^4$           & $34.26$  & $0.115$    &   $ 0.064  \pm  0.006$   &  $0.828 $   \\
\hline
$18^4$           & $11.99$  & $0.115$    &   $0.046 \pm 0.009$      & $ 0.947 $   \\
\hline
 $18^4$          & $17.13$  & $0.115$    &   $0.061 \pm 0.003$      & $0.137$  \\
\hline
 $18^4$          & $25.70$  & $0.115$    &   $0.067 \pm 0.002$      & $0.200$   \\
 \hline
$18^4$           & $34.26$  & $0.115$    &   $0.071 \pm 0.009$      & $3.670$     \\
\hline
$18^4$           & $m_d=0\ extr.$   & $0.115$    &   $0.043 \pm 0.005$         & $ 0.576$   \\
\hline
\end{tabular}
\end{center}
\caption{The magnetic dipole polarizability   of $\pi^0$, its error
and $ \chi^2$/d.o.f. obtained from  fit \eq{eq1} to the lattice data
at  $(eB)^2 \in [0,0.15]\ \Gev^4$ for various  bare quark masses,
lattice volume  $18^4$ and  lattice spacing  $0.115\ \fm$. For the
$16^4$ lattice the fit  was done at $(eB)^2 \in [0,0.2]\ \Gev^4$.}
\label{Table2}
  \end{table}

  \begin{table}[htb]
 \begin{center}
\begin{tabular}{|r|r|r|r|r|r|}
\hline
     $V_{latt}$  & $m_d$ $(\Mev)$ &$a\ (\fm)$   &$\beta_m\,(\Gev^{-3})$  & $\beta_m^h\,(\Gev^{-7})$  & $ \chi^2$/d.o.f.  \\
 \hline
    $18^4$          & $17.13$  & $0.115$    &   $0.056 \pm 0.007$      & $-0.068 \pm 0.015$     & $1.433 $  \\
 \hline
     $18^4$           & $34.26$  & $0.115$    &  $0.061 \pm 0.007 $     & $ -0.059 \pm 0.013$     & $7.622$  \\
\hline
    $18^4$           & $34.26$  & $0.095$    &  $0.064 \pm 0.009$     & $  -0.058 \pm 0.015$      & $0.673$  \\
\hline
    $20^4$           & $34.26$  & $0.115$    &  $0.069 \pm 0.007$     & $  -0.066 \pm 0.011$      & $10.846$  \\
\hline
\end{tabular}
\end{center}
\caption{The magnetic dipole polarizability and  hyperpolarizability
of $\pi^0$, its errors  and $ \chi^2$/d.o.f. obtained from the fit
\eq{eq2} to the lattice data at $(eB)^2 \in [0,1.7]\ \Gev^4$ for
various  quark masses, lattice volumes   $18^4$, $20^4$  and lattice
spacings $0.095\ \fm$, $0.115\ \fm$.} \label{Table3}
  \end{table}

\end{document}